\documentclass{article}
\usepackage{spconf,amsmath,graphicx}
\usepackage[caption=false, font=footnotesize]{subfig}% Modern replacement for subfigure package
\usepackage{paralist}% extended list environments
\usepackage{booktabs}
\title{Phase Repair for Time-Domain Convolutional Neural Networks in Music Super-Resolution}
\name{Yenan Zhang$^1$, Guilly Kolkman$^2$, Hiroshi Watanabe$^1$}
\address{$^1$Graduate School of Fundamental Science and Engineering, Waseda University, Tokyo, Japan\\
        $^2$The Faculty of Science, University of Amsterdam, Amsterdam, Netherlands}
\begin{document}
%\ninept
%
\maketitle
\begin{abstract}
Audio Super-Resolution (SR) is an important topic as low-resolution recordings are ubiquitous in daily life. 
In this paper, we focus on the music SR task, which is challenging due to the wide frequency response and dynamic range of music. 
Many models are designed in time domain to jointly process magnitude and phase of audio signals. However, prior works show that approaches using Time-Domain Convolutional Neural Network (TD-CNN) tend to produce annoying artifacts in their waveform outputs, and the cause of the artifacts is yet to be identified. 
To the best of our knowledge, this work is the first to demonstrate the artifacts in TD-CNNs are caused by the phase distortion via a subjective experiment. We further propose Time-Domain Phase Repair (TD-PR), which uses a neural vocoder pre-trained on the wide-band data to repair the phase components in the waveform outputs of TD-CNNs. 
Although the vocoder and TD-CNNs are independently trained, the proposed TD-PR obtained better mean opinion score, significantly improving the perceptual quality of TD-CNN baselines. Since the proposed TD-PR only repairs the phase components of the waveforms, the improved perceptual quality in turn indicates that phase distortion has been the cause of the annoying artifacts of TD-CNNs. 
Moreover, a single pretrained vocoder can be directly applied to arbitrary TD-CNNs without additional adaptation. Therefore, we apply TD-PR to three TD-CNNs that have different architecture and parameter amount. Consistent improvements are observed when TD-PR is applied to all three TD-CNN baselines. Audio samples are available on the demo page\footnote{https://mannmaruko.github.io/demopage/tdpr.html}.
  %
  %from 2.5kHz to 4kHz with a target bandwidth of 8kHz
  %, which shows the generalization ability of the proposed TD-PR 
\end{abstract}

\section{Introduction}\label{sec:introduction}
Audio Super-Resolution (SR), also known as bandwidth extension and bandwidth expansion, aims to predict the High-Resolution (HR) components from the Low-Resolution (LR) input audio. 
Audio SR is an important topic as LR audio is common in daily life, \textit{e.g.}, historical recordings or unprofessional-made modern recordings. 
The real-world LR recordings have a variety of bandwidth or even ambiguous bandwidth. Therefore, addressing audio SR in real world is challenging. 
Deep Neural Networks (DNNs) have become the mainstream on audio SR tasks \cite{kuleshov2017AudioUnet, hu2020PhaseAwareinterspeech, li2021realtimeSEANet, liu2022neuralvocoderisallyouneed}, but only a few works focus on the music \cite{hu2020PhaseAwareinterspeech}. 
To investigate the music SR task, we focus on solo piano recordings, as piano is a representative instrument with the most broad frequency range among other music instruments.% focus on the music SR with solo piano music.  We perform SR on solo instrument music instead of the orchestra, since the task will be simpler on single source data. Among many music instruments, we choose solo piano music as the representative, since piano has the most broad frequency range.
% compared with conventional methods

Various works have delved into the DNN-based approaches for audio SR. 
Frequency-Domain Convolutional Neural Networks (FD-CNNs) try to directly recover the HR components in the magnitude spectrogram, and generally require additional signal processing to estimate the corresponding phase information, such as Griffin-Lim algorithms \cite{hu2020PhaseAwareinterspeech} or a neural vocoder \cite{liu2022neuralvocoderisallyouneed}. 
Compared with FD methods, Time-Domain Convolutional Neural Networks (TD-CNNs) that directly learn a wave-to-wave mapping, are considered being able to avoid the phase problem on audio SR tasks \cite{hu2020PhaseAwareinterspeech}. 
However, TD-CNNs (\textit{e.g.}, AudioUNet \cite{kuleshov2017AudioUnet}) tend to produce annoying artifacts in their waveform output. To alleviate the artifacts, Lim \textit{et\;al.} proposed a time-frequency hybrid model \cite{lim2018timefrequcencynetworksforaudiosr} based on AudioUNet. Wang \textit{et\;al.} made efforts on objective function that employing the frequency domain losses \cite{wang2021towardsrobustspeechsr} during the TD-CNN's training. 
The data augmentation strategy was proposed in \cite{FilterGenerforMusicBandwithExtension} to improve the robustness of TD-CNNs. 

Although the above efforts for TD-CNN improved audio SR quality measured by objective scores, none of the above TD-CNN methods succeeds in removing the artifacts according to their open-available audio samples. 
We hypothesize that the inconsistency between objective and subjective evaluation results could have been caused by some signal components that cannot be measured by the objective metrics. We observed that phase components are not explicitly measured by typical objective metrics such as log-spectral distance. 
This observation encourages us to explore the importance of phase in audio SR tasks. In terms of up-sampling ratio, many works perform the SR on the fixed ratio (\textit{e.g.}, 2$\times$) \cite{kuleshov2017AudioUnet, hu2020PhaseAwareinterspeech}, which would be a limitation when apply these models to real world scenarios.
    
We investigate the artifacts of TD-CNNs in the following ways. 
First, we train three TD-CNNs with different architecture and parameter amount to handle LR music with various bandwidth, which is applicable to real world problems. 
We successfully reproduced the SR capability as well as the artifacts for three TD-CNN baselines. 
Second, we conduct an AB listening test which, to the best of our knowledge, is the first to demonstrate the artifacts in TD-CNNs are caused by the phase distortion via a subjective experiment. 
Last but not least, we propose the Time-Domain Phase Repair (TD-PR) method, which utilizes a vocoder pretrained on wide-band music signals to repair the distorted phase components in the waveform output of TD-CNN baselines. 
Since the vocoder and TD-CNNs are trained independently, a single pretrained vocoder can be directly applied to arbitrary TD-CNNs without additional adaptation. 
Therefore, we apply TD-PR to the aforementioned three TD-CNNs. The proposed TD-PR consistently and significantly improved the perceptual quality of all three TD-CNN baselines. 
Since TD-PR only repair the phase components of the waveforms, the improved perceptual quality in turn indicates that phase distortion has been the cause of the annoying artifacts of TD-CNNs.
    
%\url{https://smcnetwork.org/smc2024/}
%Changes for 2021 version (maintained in the 2024 version):
%\begin{itemize}
%	\item Line numbering for Review purposes (Please turn it off for Camera Ready)
%	\item Support of non latin alphabets for authors (see template for guidelines)
%	\item Support of structured names for authors (firstname, middlename, lastname, generation, phonetic variants, email, orcid)
%	\item Semistructured affiliation support (labunit, department, institution, streetaddress, city, state, postcode, country)
%	\item Explicit mapping for affiliations and authors (Based on authoblk)
%\end{itemize}

\section{Related work}
\label{sec:related_work}
Various approaches for audio SR have been developed and some of them work in Frequency Domain (FD). 
Li \textit{et\;al.} proposed an FD approach for speech SR, which consists of 2 steps \cite{li2015ADeepNeuralNetworkApproachtoSBWE}. 
The first step is mapping the magnitude components from narrow-bandwidth to wide-bandwidth by DNN. 
The second step is to estimate the corresponding phase by signal processing. Following this work, Hu \textit{et\;al.} introduced Generative Adversarial Network (GAN) into both steps and got the better performance \cite{hu2020PhaseAwareinterspeech}. 
However, training two GAN-based models is difficult due to the instability of GAN training. Furthermore, this SR system works on a fixed up-sampling ratio, which limits it's application to real world problems. Liu \textit{et\;al.} used a GAN-based neural vocoder for the second step without using GAN in the first step, which successfully performed speech SR  with the ability of handling various up-sampling ratios \cite{liu2022neuralvocoderisallyouneed}. 
It's worth pointing out that the FD approaches mentioned above requires strict matching of mel-spectrogram settings between the FD-CNN model and the neural vocoder. Therefore, some FD-CNN models trained with an unmatched mel-spectrogram settings cannot directly work with the pretrained vocoder.

Contrary to FD approaches, TD-CNNs are considered being able to avoid the phase problem in the audio SR tasks due to the direct waveform processing  \cite{hu2020PhaseAwareinterspeech}. 
AudioUNet is one of the pioneers of tackling audio SR by the TD-CNN \cite{kuleshov2017AudioUnet}. 
Tagliasacchi \textit{et\;al.} proposed SEANet \cite{Tagliasacchi2020SAENet}, a GAN-based model for speech SR. 
The generator of SEANet is a light-weight but effective TD-CNN. In this paper, We utilized the generator of SEANet to music SR as one of our baselines. 
Defossez \textit{et\;al.} proposed TD-CNN model named Demucs, which is a large model with over 130M parameters and is initially designed to address music source separation \cite{defossez2019demucs}. Considering the fact that Demucs has shown strong performance in tasks besides source separation \cite{su2021bandwidthextensionisallyouneed}, we utilize the Demucs model in the SR task in this paper. 
To the best of our knowledge, this is the first time to apply Demucs to the music SR task.
    
The mel-to-wave transform is commonly addressed by the neural vocoder. TFGAN is a light-weight vocoder \cite{tian2020tfgan} and has been applied to speech SR task \cite{liu2022neuralvocoderisallyouneed}.
	
\section{Proposed method}\label{sec:proposed_method}
%\subsection{Poor Phase Is The Cause of The Artifact}\label{AA}
    %As mentioned in section \ref{sec:introduction}, a variety of works have tried to alleviate the artifact of TD-CNN approaches, including modifying the network architecture \cite{lim2018timefrequcencynetworksforaudiosr}, using loss functions that in both frequency and time domain \cite{wang2021towardsrobustspeechsr}. Although objective metrics measured by magnitude similarity indicate the effectiveness of the above TD-CNN methods, none of them succeeds in removing the artifacts. Considering phase components are not explicitly measured by the standard objective metric for the audio SR task, the inconsistency between objective and subjective results could have been caused by components that cannot be measured by the objective metric. This observation encourages us to explore the importance of phase in audio SR tasks.
    %Instead, we observed that the artifact exists in the TD-CNN's result even when the corresponding spectrogram has been recovered to a very good quality. As there is no obvious artifact in magnitude, we consider phase, the other component of STFT spectrogram, may be the cause of the artifact. We confirm it through the listening test I (\textit{i.e.},an AB test) in which we replace the phase of the TD-CNN output with the phase of the corresponding ground truth. The result in section \ref{results} implies the artifact of TD-CNN approaches in audio SR is caused by the phase distortion. To the best of our knowledge, this is the first time to demonstrate the phase issue via a subjective test.
    
\begin{figure}[tb] 
    \centering
    \includegraphics[width=\linewidth]{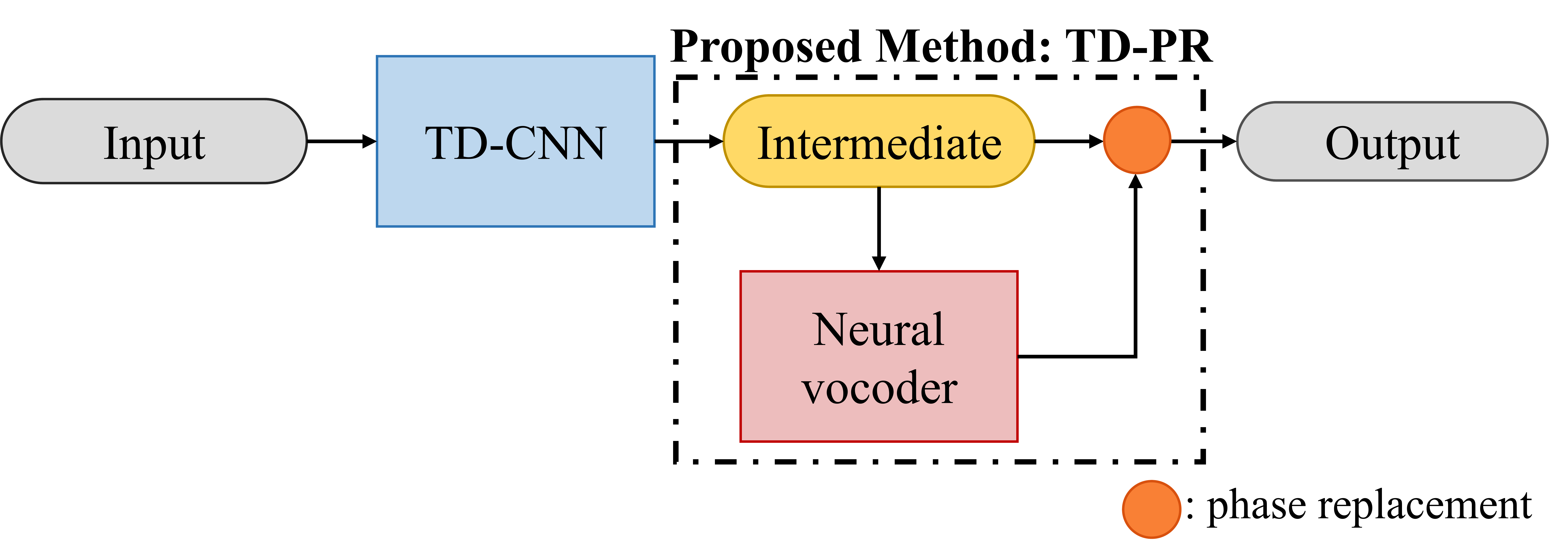}
    \caption{Overview of the proposed TD-PR: The TD-CNN is trained to perform super-resolution for various narrow-band inputs. The neural vocoder takes only the magnitude of the TD-CNN's output as input, and re-synthesizes another waveform that contains repaired phase components. Then, the distorted phase components in TD-CNN's output is replaced by that from the vocoder.}
    \label{fig:overview}
\end{figure}

\subsection{Time-Domain Phase Repair}
\label{subsec:time_domain_phase_repair}

In order to alleviate the artifacts caused by distorted phase components, we propose Time-Domain Phase Repair (TD-PR). The TD-PR framework consists of two separately pretrained DNN modules and a phase replacement operation. 
    
The overview of the proposed method is shown in Fig. \ref{fig:overview}. Specifically, the TD-PR pipeline involves the following steps. First, we train a TD-CNN to perform music SR. 
To handle LR music with various bandwidths which is common in real world, we apply a simulation pipeline to HR music data to get the corresponding LR version. With the simulated pseudo paired data, the training of TD-CNN for music SR is made possible. 
Details of the simulation pipeline and training objectives are explained in the succeeding section. 

Second, we pretrain a neural vocoder on the unprocessed HR music data. Since a neural vocoder can generate realistic waveform signals with only the magnitude input, it can be inferred that a vocoder can generate realistic phase components that are coherent with the input magnitude components. This inspires us to utilize a neural vocoder to repair distorted phase.
    %We conduct an AB listening test which, to the best of our knowledge, is the first to demonstrate the artifact in TD-CNNs is caused by the phase distortion via a subjective test. 
    
Last, we introduce TD-PR to repair the phase components of the SR output from a TD-CNN. Assume that the TD-CNN is applied to an LR music source and generated an intermediate waveform, which is then decomposed into magnitude and phase comonents by Short-Time Fourier Transform (STFT). We empirically decided to use an STFT of 1024-point hann window and 256 hop length for a sampling rate of 16 kHz. The neural vocoder takes only the magnitude of the TD-CNN's output as input, and re-synthesizes another waveform that contains repaired phase components. Then, the distorted phase components in TD-CNN’s output is replaced by that from the vocoder, and a phase-repaired waveform output is produced by inverse STFT. 
    
According to the above description, the vocoder and TD-CNNs are trained independently, which means a single pretrained vocoder can be directly applied to arbitrary TD-CNNs without additional adaptation, making the method flexible. Last but not least, it is worth noting that since TD-PR only repair the phase components of the waveforms, the improved perceptual quality in turn indicates that phase distortion has been the cause of the annoying artifacts of TD-CNNs.
    
   %An example of a set of spectrograms, including LR input, ground truth, the result of SEANet, SEANet w/ TD-PR (proposed), AudioUNet, AudioUNet w/ TD-PR (proposed), Demcus, Demucs w/ TD-PR (proposed) is shown in Fig. \ref{fig:spectrograms}. 
   %One of the merits of our proposed method is that it can be directly used for general TD-CNN models without parameter adjustment, since the output of TD-CNNs is waveform and it can be converted to mel-spectrogram in arbitrary settings, no matter what mel-spectrogram settings the neural vocoder requires. This is different from the FD method, which requires strictly matched configurations as mentioned in section \ref{related work}. 
   %The TD-PR module in the proposed method is trained by inputting clean and wide-bandwidth music, so the proposed method has the potential to be applied to TD-CNNs that are trained for other tasks, such as music enhancement.

\subsection{Simulation Pipeline} \label{subsec:simulation_pipeline}

The design of simulation pipeline has been shown important to the performance and robustness of audio SR models \cite{wang2021towardsrobustspeechsr, FilterGenerforMusicBandwithExtension}. Therefore, we decided to follow the principles in \cite{wang2021towardsrobustspeechsr, FilterGenerforMusicBandwithExtension}. Specifically, we simulate each LR input by randomly choosing a low-pass filter from 7 low-pass filters, including Butterworth, Chebyshev type 1, Chebyshev type 2, Elliptic, Bessel, subsampling (\textit{i.e.}, resample\_poly in scipy), STFT filter (\textit{i.e.}, replacing the high frequency components with zero elements) with the filter order randomly selected from 6 to 10. We used the implementation of low-pass filters provided by Liu \textit{et\; al.}$\footnote{https://github.com/haoheliu/ssr\_eval}$ \cite{liu2022neuralvocoderisallyouneed}. 

Since 3 kHz is analyzed to be the typical bandwidth of real historical recordings \cite{moliner2022behmgan}, we uniformly sample an LR bandwidth between 2.5 kHz and 4 kHz. We don't consider 2 kHz because we found this bandwidth will filter out a part of melody, which is not common in real recordings.
The low-pass filtering is conducted on-the-fly during training. 
    %The range of the LR bandwidth is determined as follows: lower bound is set to be 2.5kHz, as 2kHz is so low that some music notes disappear at this bandwidth. Upper bound is 4kHz, as this is the common case of the $2\times$ up-sampling ratio for SR \cite{hu2020PhaseAwareinterspeech,liu2022neuralvocoderisallyouneed}. The target bandwidth is 8kHz, the maximum frequency of 16kHz sampling rate.
\subsection{Loss Function} \label{subsec:loss_function}
\vspace{1mm}
Inspired by \cite{wang2021towardsrobustspeechsr}, we perform cross-domain loss to guide TD-CNNs to capture features in both time and frequency domains. The loss function (denoted as $L$) is comprised of two parts, multi-resolution STFT loss ($L_{\textrm{MRSTFT}}$) \cite{yamamoto2020parallelwavegan} and multi-resolution wave loss($L_{\textrm{MRwave}}$) which is similar to $L_{\textrm{MRSTFT}}$. The loss function is defined as below:
    \vspace{-1mm}
    \begin{equation}
    L = L_{\textrm{MRSTFT}} + \lambda L_{\textrm{MRwave}}, \label{eq}
    \end{equation}
    where $\lambda$ denotes the hyperparameter balancing the two loss terms. In our case, we empirically set $\lambda = 1000$ to balance the weights between two losses. 
    
    The definition of $L_{\textrm{MRSTFT}}$ and $L_{\textrm{MRwave}}$ are shown as follows:
    \vspace{-1mm}
    \begin{equation}    
    L_{\textrm{MRSTFT}} = \frac{1}{M}\sum_{m=1}^{M}L^{(m)}_{\textrm{STFT}}(y, \hat{y}), \label{eq}
    \end{equation}
    \vspace{-2.5mm}
    \begin{equation}
    L_{\textrm{MRwave}} = \frac{1}{N}\sum_{n=1}^{N}L^{(n)}_{\textrm{wave}}(y, \hat{y}), \label{eq}
    \end{equation}
    \vspace{-1mm}
    where $y$ and $\hat{y}$ denote the ground truth and generated sample respectively. $M$ denotes the number of STFT losses with different analysis parameters (\textit{i.e.}, FFT size = [512, 1024, 2048]; hop size = [256, 512, 1024]; window size = [512, 1024, 2048]). We used the implementation of $L_{\textrm{MRSTFT}}$ from \cite{steinmetz2020auraloss}. $N$ denotes the number of wave losses with different sampling rate (\textit{i.e.}, original sampling rate, $2\times$down sampling rate, $4\times$down sampling rate). 
    
     $L_{\textrm{wave}}$ is defined as follows:
    %\begin{equation}
    %L_{\textrm{STFT}}(y, \hat{y}) = L_{\textrm{SC}}(y, \hat{y}) + L_{\textrm{mag}}(y, \hat{y})\label{eq}
    %\end{equation}
    \begin{equation}
     L_{\textrm{wave}}(y, \hat{y}) = \frac{1}{P}\Vert\,y -\hat{y}\,\Vert_1, \label{eq}
    \end{equation}
    where $P$ denotes the number of wave samples and $\Vert\,\cdot\,\Vert_1$ denotes the L1 norms.
    %$L_{\textrm{SC}}$ and $L_{\textrm{mag}}$ denote spectral convergence and log-STFT magnitude loss respectively, which are defined as follows:
    %\begin{equation}
     %L_{\textrm{SC}}(y, \hat{y}) = \frac{\Vert\,\lvert\textrm{STFT}(y)\rvert-\lvert\textrm{STFT}(\hat{y})\rvert\,\Vert_F}{\Vert\,\lvert\textrm{STFT}(y)\rvert\,\Vert_F}\label{eq}
    %\end{equation}
    %\begin{equation}
     %L_{\textrm{mag}}(y, \hat{y}) = \frac{1}{Q}\Vert\,\textrm{log}\lvert\textrm{STFT}(y)\rvert -\textrm{log}\lvert\textrm{STFT}(\hat{y})\rvert\,\Vert_1\label{eq}
    %\end{equation}
    %where $\Vert\,\cdot\,\Vert_F$ denotes the Frobenius norm; $\lvert\,\textrm{STFT}(\cdot)\,\rvert_F$ and $Q$ denote the STFT magnitudes and number of elements in the magnitude respectively.
    
    \section{Experiments}
    \subsection{Dataset And Implementation} \label{sec:dataset_implementation}
    \vspace{1mm}
    We train and evaluate our model on the MAESTRO dataset \cite{Hawthorne2018MAESTRO}. It's composed of about 200 hours of high-quality classical piano recordings in waveform. Although these recordings have the sampling rate of 44.1 kHz or 48 kHz, we empirically found that 16 kHz is high enough for the piano solo. Hence, we decided to perform music SR on the target bandwidth of 8 kHz, \textit{i.e.}, a target sampling rate 16 kHz. We used the official split of MAESTRO for training, validation and testing. We cut all of the waveform into 30-second short clips for efficient training. %Besides, working at 16 kHz also helps to reduce the computational cost

    To implement the proposed TD-PR framework, we trained a TFGAN \cite{tian2020tfgan} from scratch on MAESTRO train set by using an unofficial implementation$\footnote{https://github.com/rishikksh20/TFGAN}$. We follow the original settings, except resetting the sampling rate to 16 kHz, and trained it for 1M iterations. 
    
    Since TD-PR is feasible for arbitrary TD-CNNs with a single pretrained neural vocoder as mentioned in Sec.\ref{subsec:time_domain_phase_repair}, we evaluate TD-PR with three representative TD-CNN models as baselines: AudioUNet \cite{kuleshov2017AudioUnet}, Demucs \cite{defossez2019demucs} and SEANet generator \cite{Tagliasacchi2020SAENet}. 
    We trained them from scratch with the loss function mentioned in Sec.\ref{subsec:loss_function} by applying the simulation pipline in Sec.\ref{subsec:simulation_pipeline} to the dataset. We used the Pytorch implementation of AudioUNet$\footnote{https://github.com/serkansulun/deep-music-enhancer}$ and Demucs$\footnote{https://github.com/facebookresearch/demucs/tree/v2}$. We implemented the SEANet generator by ourselves. We used an Adam optimizer and the initial learning rate 0.0001 to optimize each TD-CNN model for 200 epochs with the batch size of 12 and the input duration of 5s. 
    
    \subsection{Investigation on The Effectiveness of Ground Truth Phase Components}
    \label{subsec:listening_test_i}
    \vspace{1mm}
    Before delving into the evaluation of TD-PR, we present a preliminary study to show the impact of phase on the artifacts issue of TD-CNN models. In this study, we used SEANet as the TD-CNN baseline, and replaced the phase of the TD-CNN output with the phase of the corresponding HR music (ground truth that is not available in real world applications), \textit{i.e.}, TD-CNN w/ GT-phase. 
    
    We then conducted an AB listening test, in which we asked participants to choose the one containing fewer artifacts between the TD-CNN baseline and TD-CNN w/ GT-phase. We selected eleven music pieces for the listening test which cover different periods and styles of different musicians from the MAESTRO test set. Eleven audio pairs are presented in the AB test, in which one pair is for practice and the left ten pairs are for evaluation. Each clip is cut into the duration of 5s. We also regularized the volume of all the samples by Audacity$\footnote{https://www.audacityteam.org/}$. The input bandwidth for this listening test is set to 3 kHz, as it is analyzed to be the typical bandwidth of historical recordings \cite{moliner2022behmgan}.
\vspace{2mm}
\subsection{Comparison between TD-PR And TD-CNN Baselines}
\label{subsec:listening_test_ii}
TD-PR is proposed to improve the perceptual quality of TD-CNN baselines via phase repair. We evaluate the proposed TD-PR from both objective and subjective aspects. In terms of the objective evaluation, we use the Log-Spectral Distance (LSD) as the metric, which has been widely used in audio SR tasks \cite{kuleshov2017AudioUnet, hu2020PhaseAwareinterspeech, liu2022neuralvocoderisallyouneed}. LSD is designed as:
    \begin{equation}
    LSD = \frac{1}{L}\sum_{l=1}^{L}\sqrt{
        \frac{1}{F}{\sum_{f=1}^{F}\left(
            \log\lvert{Y_{l,f}\rvert^{2}-\log\lvert\hat{Y}_{l,f}\rvert^{2}}
        \right)^{2}}
        }, \label{eq}
    \end{equation}
    where $Y_{l,f}$ and $\hat{Y}_{l,f}$ are the ground truth and the estimated magnitude via STFT at $l$-th time step $(l = 1, ..., L)$ and $f$-th frequency bin $(k = 1, ..., F)$, respectively.
    
The subjective evaluation aims at collecting Mean Opinion Score (MOS) from participants to compare the perceptual quality across the input LR music, TD-CNN baseline, TD-CNN w/ TD-PR and ground truth HR music. MOS is commonly used in audio SR tasks to represent the perceptual quality \cite{liu2022neuralvocoderisallyouneed, su2021bandwidthextensionisallyouneed}. Participants are asked to rate audio samples according to the similarity with the reference audio, \textit{i.e.}, the groud truth HR music. The range of MOS in our work is set from 1 to 5, where 5 denotes excellent quality (\textit{i.e.}, is closest to the reference) and 1 denotes bad quality. To avoid auditory fatigue caused by giving too many samples to participants, we evaluate the three TD-CNN models separately in three independent listening tests, which means the MOS values across different tests cannot be directly compared. The same eleven music pieces and pre-processing as in the preliminary AB test are used.

\section{Results and Discussion} 
\label{sec:results}
\subsection{Impact of Ground Truth Phase Components}
\label{subsec:result_ABtest}
\begin{figure}[t] 
    \centering
    \includegraphics[width=\linewidth]{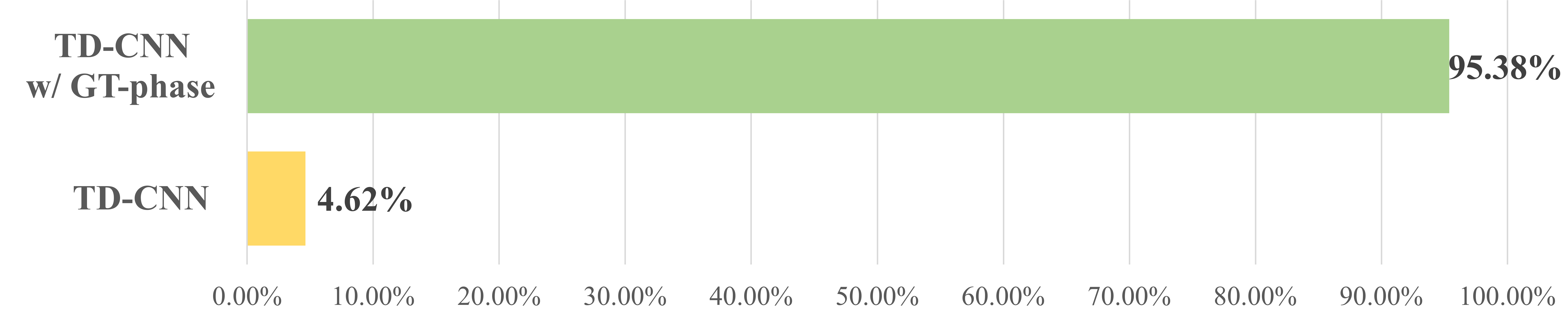}
    \caption{Results of the preliminary AB listening test: 95.38\% of the TD-CNN w/ GT-phase is voted to have fewer artifacts.}
    \label{fig:abtestresults}
\end{figure}

The preference of the AB listening test between TD-CNN baseline and TD-CNN w/ GT-phase described in Sec. \ref{subsec:listening_test_i} is shown in Fig. \ref{fig:abtestresults}. TD-CNN w/ GT-phase is voted to have fewer artifacts with a large margin (95.38\% vs 4.62\%). Therefore, we concluded that the artifacts in TD-CNN approaches for audio SR tasks is caused by the phase distortion, and the distortion can be repaired by replacing the distorted phase with a more realistic one.
    
\subsection{Results on TD-PR}
\label{subsec:result_tdpr}

\begin{figure}[h] 
    \centering
    \includegraphics[width=\linewidth]{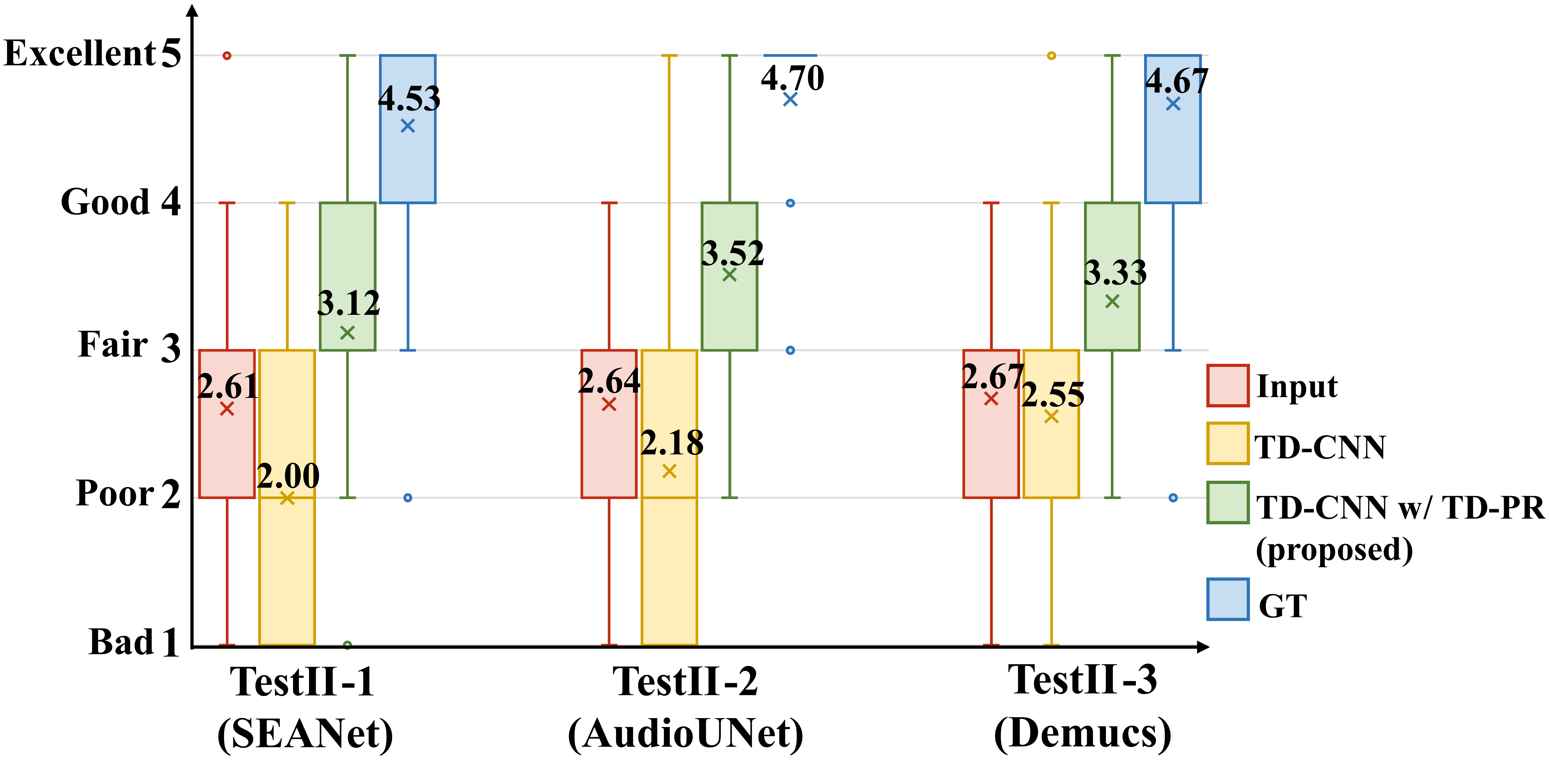}
    \caption{Results of MOS listening test: The box plot of the ratings across input, TD-CNN, TD-PR and GT. TD-PR is applied to three different TD-CNN baselines.}
    \label{fig:mosboxplot}
\end{figure}

We conducted the MOS listening test described in Sec.\ref{subsec:listening_test_ii}. The box plot of the MOS test results and the corresponding average for each method are shown in Fig. \ref{fig:mosboxplot}. 
First, the proposed TD-PR obtained better MOS scores than all three TD-CNN baselines by a large margin, \textit{e.g.}, the proposed TD-PR has higher boxes, and higher average MOS scores of 1.12 (SEANet), 1.34 (AudioUNet), 0.78 (Demucs), revealing that the TD-PR improved the perceptual quality of TD-CNN baselines significantly. Successfully improving three different baselines with a single pretrained vocoder indicates the flexibility of the proposed TD-PR method. 

    Looking at the average MOS scores between input LR music and TD-CNN baselines, it is obversed that TD-CNN baselines obtained lower MOS than the LR input by the deterioration of -0.61 (SEANet), -0.46 (AudioUNet), -0.12 (Demucs). This indicates that the artifacts in TD-CNNs severely harmed the perceptual quality. However, we will show later that TD-CNN baselines obtained better LSD scores (objective metric) than the LR input, showing that LSD is not a reliable metric to evaluate audio SR and perceptual quality. 
\begin{table}[t]
    \caption{LSD results with different input bandwidth and parameter amount of different models.}
    %\vspace{-5mm}
    \begin{center}
    \resizebox{\linewidth}{!}{
    \begin{tabular}{lcccc|c|c}
    \toprule
    %\toprule%[2pt]
      & 2.5kHz & 3kHz & 3.5kHz & 4kHz & AVG & Parameter \\
    \midrule
    Input & 2.43 & 2.19 & 1.97 & 1.78 & 2.09 & - \\
    \midrule
    SEANet & 0.89 & 0.78 & 0.72 & 0.68 & 0.77 & 11M\\
    %\midrule
    SEANet w/ TD-PR(proposed)  & 0.94 & 0.86 & 0.82 & 0.80 & 0.86 & 11+6M\\
    \midrule
    %\midrule
    AudioUNet & 0.83 & \textbf{0.74} & 0.69 & 0.66 & 0.73 & 56M\\
    %\midrule
    AudioUNet w/ TD-PR(proposed) & 0.89 & 0.82 & 0.79 & 0.77 & 0.82 & 56+6M\\
    \midrule
    %\midrule
    Demucs & \textbf{0.82} & \textbf{0.74} & \textbf{0.68} & \textbf{0.64} & \textbf{0.72} & 134M\\
    %\midrule
    Demucs w/ TD-PR(proposed) & 0.89 & 0.83 & 0.79 & 0.77 & 0.82 & 134+6M\\
    %\midrule
    \midrule
    Ground truth & 0 & 0 & 0 & 0 & 0 & -\\
    \bottomrule
    %\bottomrule%[2pt]
    \end{tabular}
    }
    \end{center}
    \label{table:lsd}
    %\vspace{-0.05em}
\end{table}    
    In terms of the gap of the average MOS between input and TD-CNN baselines, Demucs showed the smallest gap to the input, which implies that Demucs is the strongest among the three baselines. This observation is also in consistency with its largest parameter amount.

    The LSD scores on 4 representative LR bandwidth (2.5 kHz, 3 kHz, 3.5 kHz, 4 kHz) is shown in Table \ref{table:lsd}. Note that the proposed method can deal with any bandwidth between 2.5 kHz and 4 kHz. The results show that both TD-PR and their TD-CNN baselines got much lower LSD than LR input, meaning that music SR is successfully achieved. Although the proposed method got sightly worse LSD scores than the baselines, we argue this is trivial, because the aforementioned MOS listening test revealed a significant gap in perceptual quality between TD-PR and baselines. Although LSD can well reflect how well the high frequency magnitude is recovered in each model, it can't reflect the degree of the phase artifact and has been observed not highly correlated with perceptual audio quality in previous literature \cite{liu2022neuralvocoderisallyouneed}.

\subsection{Qualitative Evaluation of TD-PR}
\label{subsec:qualitative}
\begin{figure}[t] 
\centering
\includegraphics[width=\linewidth,scale=1.00]{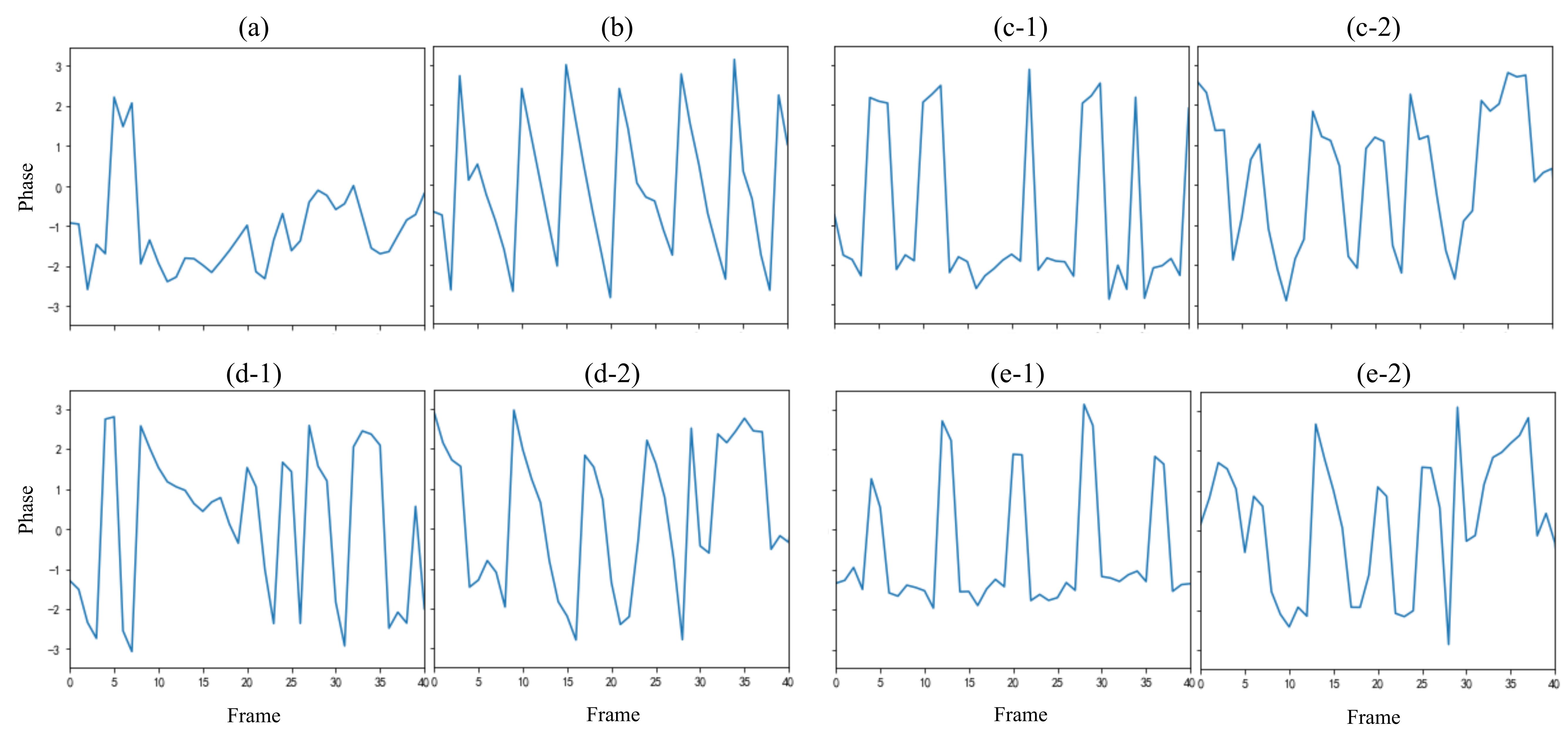}
    %\vspace{-6mm}
    \caption{Visualization of a set of phase spectrograms: (a) low-resolution input; (b) ground truth; (c-1) SEANet; (c-2) SEANet w/ TD-PR (proposed); (d-1) AudioUNet; (d-2) AudioUNet w/ TD-PR (proposed); (e-2) Demcus; (e-2) Demucs w/ TD-PR (proposed).}
    \label{fig:phase spectrum}
\end{figure}

\begin{figure}[t] 
    \centering
    \includegraphics[width=\linewidth,scale=1.00]{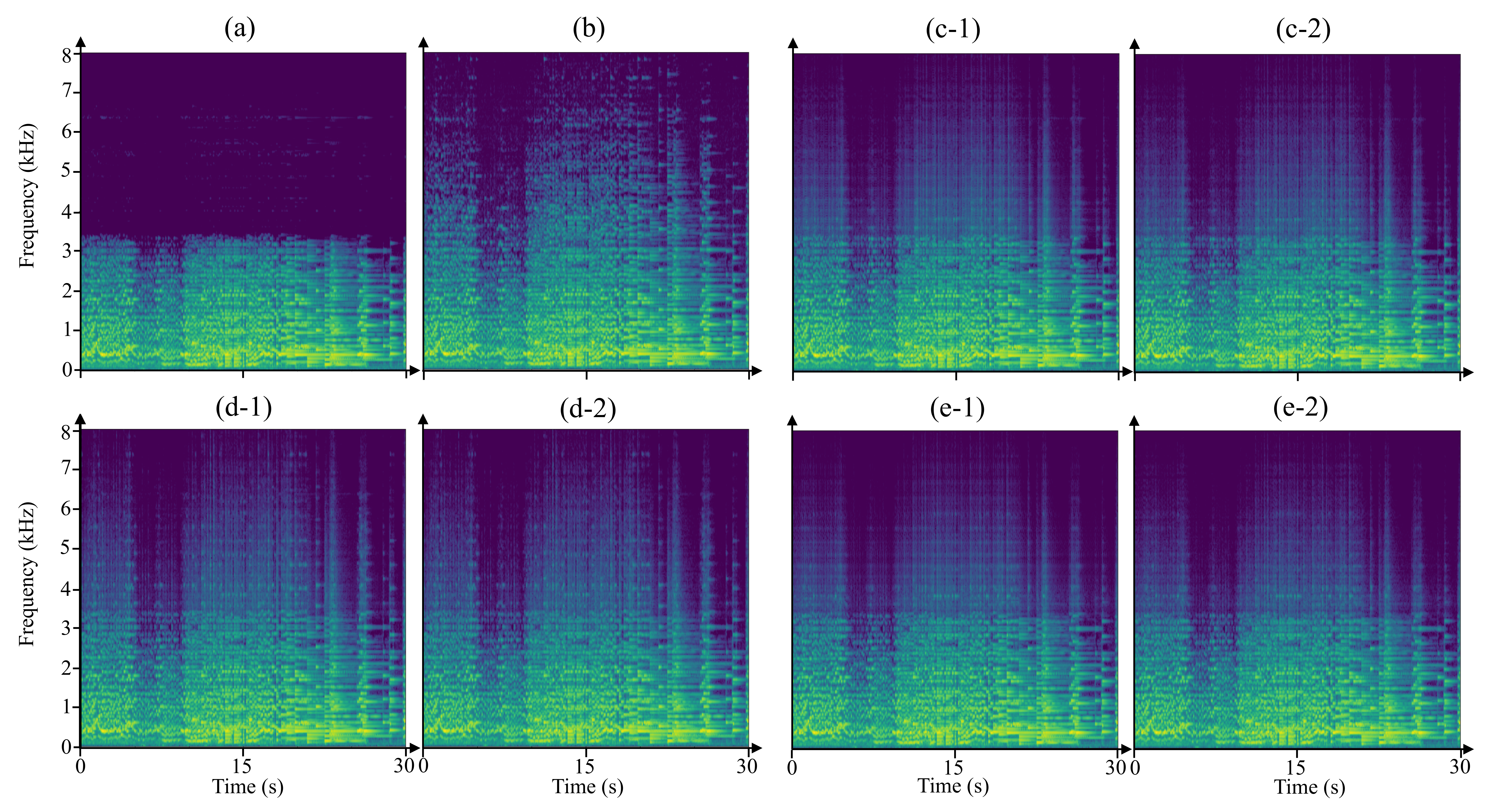}
    \caption{Visualization of a set of magnitude spectrograms: (a) low-resolution input; (b) ground truth; (c-1) SEANet; (c-2) SEANet w/ TD-PR (proposed); (d-1) AudioUNet: (d-2) AudioUNet w/ TD-PR (proposed); (e-2) Demcus; (e-2) Demucs w/ TD-PR (proposed).}
    \label{fig:magnitude spectrum}
\end{figure}

    We visualize a part of phase spectrograms in Fig. \ref{fig:phase spectrum} and their corresponding magnitude spectrograms in Fig. \ref{fig:magnitude spectrum} to qualitatively evaluate the proposed TD-PR method. The visualizations include the spectrograms of LR input, ground truth, three TD-CNN baselines and their corresponding TD-PR outputs. For a clear view in Fig. \ref{fig:phase spectrum}, we plot only the phase of a single frequency bin for the first 40 time frames of an audio sample, as the phase spectrogram across multiple frequency bins is difficult to understand. The visualizations reveal that the proposed TD-PR successfully produced a phase distribution that is closer to ground truth's compared to TD-CNN baselines. Meanwhile, as TD-PR only repairs the phase components, we cannot observe significant differences in magnitude spectrograms shown in Fig. \ref{fig:magnitude spectrum}. Nevertheless, perceptual quality is improved significantly by TD-PR. The visualizations again validate that phase distortion has been the cause of the annoying artifacts in TD-CNNs.
    %Therefore the collected MOS can reflect the overall perceptual quality of each model. 

\section{Conclusion}
    In this research of music Super-Resolution (SR), we delved into Time-Domain Convolutional Neural Networks (TD-CNNs), trying to identify the cause of the annoying artifacts and improve TD-CNNs' perceptual quality by alleviating the artifacts. To the best of our knowledge, this work is the first to demonstrate the artifacts in TD-CNNs is caused by the phase distortion via a subjective experiment. We further propose Time-Domain Phase Repair (TD-PR), which uses a neural vocoder pretrained on the wide-band data to repair the phase components in the waveform output of TD-CNNs. The proposed TD-PR achieved better mean opinion score, significantly improving the perceptual quality of TD-CNN baselines. Moreover, a single pretrained vocoder can be directly applied to arbitrary TD-CNNs without additional adaptation. Since the proposed TD-PR only repairs the phase components of waveform, the improved perceptual quality in turn indicates that phase distortion has been the cause of the annoying artifacts of TD-CNNs. The findings and comprehensive evaluations presented in this work offer a new perspective for the future improvement of audio super-resolution algorithms.

%%%%%%%%%%%%%%%%%%%%%%%%%%%%%%%%%%%%%%%%%%%%%%%%%%%%%%%%%%%%%%%%%%%%%%%%%%%%%
	%bibliography here
\bibliographystyle{IEEEbib}
\bibliography{tdpr}

\end{document}